%% file: eprint_dpf2015.tex
\documentclass[12pt]{article}
\usepackage{graphicx}
\usepackage{hyperref}
\usepackage{bbm}
\usepackage{amsfonts}
\usepackage{mathrsfs}

\newcommand\pubnumber{DPF2015-373}
\newcommand\pubdate{\today}

\def\napoli{Michigan Center for Theoretical Physics\\
University of Michigan, Ann Arbor, MI 48109, USA}
\def\support{\footnote{Work supported by the Department of Energy under grants DE-SC0007879 and DE-SC0011719.}}

\def\Title#1{\begin{center} {\Large #1 } \end{center}}
\def\Author#1{\begin{center}{ \sc #1} \end{center}}
\def\Address#1{\begin{center}{ \it #1} \end{center}}

\newcommand\pubblock{\rightline{\begin{tabular}{l} \pubnumber\\
         \pubdate  \end{tabular}}}
\newenvironment{Abstract}{\begin{quotation}  }{\end{quotation}}
\newenvironment{Presented}{\begin{quotation} \begin{center} 
             PRESENTED BY BIBHUSHAN SHAKYA AT\end{center}\bigskip 
      \begin{center}\begin{large}}{\end{large}\end{center} \end{quotation}}

\newcommand{\pv}{\langle\phi\rangle}
\newcommand{\hv}{\langle H_u^0\rangle}

\input econfmacros.tex
\begin{document}
\begin{titlepage}
\pubblock

\vfill
\Title{Hints of a PeV Scale Supersymmetric Neutrino Sector:\\ 
\vskip3mm
Neutrino Masses, Dark Matter, keV and PeV Anomalies}
\vfill
\Author{ Bibhushan Shakya, Samuel B. Roland, James D. Wells\support}
\Address{\napoli}
\vfill
\begin{Abstract}

Neutrino masses and light (keV-GeV) sterile neutrinos can arise naturally via a modified, low energy seesaw mechanism if the right-handed neutrinos are charged under a new symmetry broken by a PeV scale vacuum expectation value, presumably tied to supersymmetry breaking. The additional field content also allows for freeze-in production of sterile neutrino dark matter. This framework can accommodate the recently observed 3.5 keV X-ray line, while a straightforward extension of the framework, using the new symmetry and the PeV energy scale, can explain the PeV energy neutrino events at IceCube. Together, these can therefore be taken as hints of the existence of a PeV scale supersymmetric neutrino sector. 
\end{Abstract}
\vfill
\begin{Presented}
DPF 2015\\
The Meeting of the American Physical Society\\
Division of Particles and Fields\\
Ann Arbor, Michigan, August 4--8, 2015\\
\end{Presented}
\vfill
\end{titlepage}
\def\thefootnote{\fnsymbol{footnote}}
\setcounter{footnote}{0}

\section{Introduction}

The seesaw mechanism as the explanation of the small neutrino masses motivates the existence of right-handed, sterile neutrinos. While the mass scale of such sterile neutrinos could lie anywhere from the eV to the GUT scale, the widely studied Neutrino Minimal Standard Model ($\nu$MSM) \cite{numsm} -- where the lightest sterile neutrino can be dark matter, and oscillations of the heavier sterile neutrinos can account for the baryon asymmetry of the Universe --  provides phenomenological motivations for them to be light (keV-GeV). While such phenomenological aspects make the $\nu$MSM an attractive model to study, the model contains the following wanting features: 
\begin{itemize}
\item The keV-GeV Majorana mass scale of the right-handed neutrinos is put in by hand, but the most natural choice for such a parameter is the cutoff of the theory, such as the GUT or Planck scale.
\item The model requires extremely unnatural choices of couplings in the Dirac mass terms, of order $10^{-7}$, to explain the observed neutrino oscillations through the seesaw mechanism. 
\item While the relic abundance of keV sterile neutrino dark matter is built up from the Dodelson-Widrow (DW) mechanism \cite{DW} in the $\nu$MSM, this is now ruled out by a combination of X-ray and Lyman-$\alpha$ bounds if it accounts for all of dark matter, necessitating a different dark matter production mechanism. 
\end{itemize} 
These hint at additional underlying structures in the theory that can realize these features in a more natural manner. This line of inquiry turns out to be very promising; as this note will demonstrate, the aforementioned shortcomings are readily solved if the right-handed neutrinos are charged under a new symmetry that is broken by a PeV scale vacuum expectation value (vev). As further motivation for the framework to be discussed, interesting connections can also be drawn with intermediate (PeV) scale supersymmetry and two recent observational anomalies: the 3.5 keV X-ray line signal \cite{35kev} and the PeV energy neutrinos observed by IceCube \cite{Aartsen:2014gkd}. The reader is referred to the original papers \cite{Roland:2015yoa} and \cite{Roland:2014vba} for details. 

\section{The Model}

As in the $\nu$MSM, the neutrino sector is extended by three right-handed sterile neutrinos $N_i$. To suppress the Majorana masses, which would naturally be at the GUT or Planck scale, it is assumed that these fields are not pure singlets, but charged under some new symmetry of nature -- for concreteness, a $U(1)'$.  This immediately forbids the traditional seesaw mechanism, but higher dimensional operators involving the SM and $N_i$ fields can be obtained by coupling the $N_i$ to other fields charged under the $U(1)'$. We introduce an exotic field $\phi$ that carries the opposite charge under $U(1)'$.

Working in a supersymmetric framework, we thus introduce three chiral supermultiplets $\mathcal{N}_i$ for the sterile neutrinos and a chiral supermultiplet $\Phi$, whose spin $(0,~1/2)$ components are labelled $(\tilde{N}_i,N_i)$ and $(\phi, \psi_\phi)$ respectively. These result in the following higher dimensional operators in the superpotential:
\begin{equation}
\label{eq:newterms}
W\supset\frac{y}{M_*} L H_u \mathcal{N}\Phi+\frac{x}{M_*}\mathcal{N}\mathcal{N}\Phi\Phi .
\end{equation}
Here $x$ and $y$ are dimensionless $\mathcal{O}(1)$ couplings (flavor structure neglected for now), and $M_*$ is the scale at which this effective theory needs to be UV completed with new physics, such as the scale of grand unification $M_{GUT}$ or the Planck scale $M_P$. If the scalar $\phi$ obtains a vev at the PeV scale, this breaks the $U(1)'$ and (after $H_u$ also acquires a vev) the following Dirac and Majorana mass terms are generated in the neutrino sector
\begin{equation}
m_D=\frac{y \pv\hv}{M_*},~~~~~m_M=\frac{x \pv^2}{M_*}.
\end{equation}
This resulting modified seesaw mechanism produces the following sterile and active neutrino mass scales:
\begin{equation}
\label{Mas}
m_s = m_M=\frac{x \pv^2}{M_*} \label{Ms},~~~m_a = \frac{m_D^2}{m_M}=\frac{y^2 \hv^2}{x M_*}.
\end{equation}

\begin{figure}[b!]
\centering
\includegraphics[height=2.3in]{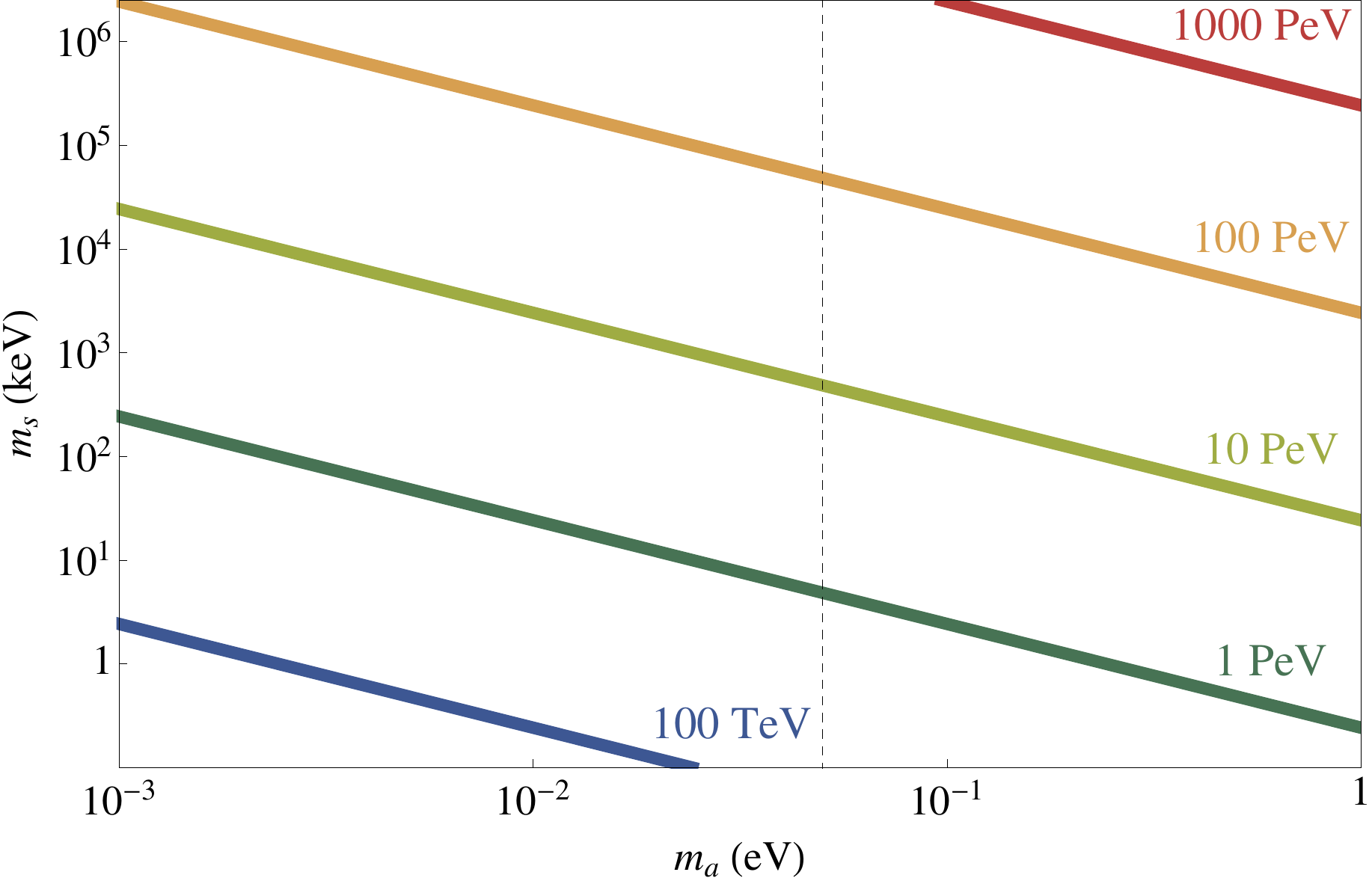}
\caption{Active and sterile neutrino mass scales for various choices of $y\langle\phi\rangle$, with $M_*\,=\,M_{GUT}$, tan$\beta=2$, and $0.001\,\textless\, x \,\textless\, 2$. The dashed vertical line denotes $m_a=0.05$\,eV, as dictated by atmospheric oscillation data $\Delta m_{atm}^2=2.3\times10^{-3}$ eV$^2$.}
\label{fig:scalology}
\end{figure}

Figure\,\ref{fig:scalology} shows possible active-sterile mass scale combinations that result from this framework with $M_*\!\!=\!\!M_{GUT}(=\!\!\!10^{16}\, {\rm GeV})$, tan$\beta\!\! =\!\!2 ~(\hv\!\!=\!\!155.6$ GeV), and $0.001\,\textless\, x \,\textless\, 2$ for various values of $y \pv$. This suggests that the desired mass scales can emerge naturally in this framework (with predominantly $\mathcal{O}(1)$ couplings) if $\langle\phi\rangle$ is at the PeV scale.

It is interesting to note that the PeV scale could be a reasonable energy scale for supersymmetry, given the null results of extensive searches for weak scale supersymmetry on various fronts. Indeed, the Higgs boson mass of $125$ GeV is consistent with PeV scale superpartners (for small tan$\beta$), and such high scale supersymmetry has been motivated for several other reasons \cite{pevsusy}. In this case, the vev of the scalar field that gives rise to masses in the neutrino sector could be tied to the mechanism for supersymmetry breaking. 

%%%%%%%%%%%%%%%%%%%%%%%%%%%%%%%%%%%%%%%%%%%%%%%%%%%%%%%%%%%%%%%%%%%%%%%%%%%

\section{Sterile Neutrino Dark Matter from Freeze-in}

We denote the sterile neutrino dark matter candidate by $N_1$. In the $\nu$MSM, $N_1$ is produced through the Dodelson-Widrow mechanism, an inevitable consequence of mixing with the active neutrinos, with abundance 
\begin{equation}
\Omega_{N_i} \sim0.2\left(\frac{\rm{sin}^2 \theta}{3\times 10^{-9}}\right)\left(\frac{m_s}{3\,\rm{keV}}\right)^{1.8}.
\end{equation}
sin$^2\theta\approx m_a/m_s$ is the mixing angle with the active neutrinos. A combination of X-ray bounds and Lyman-$\alpha$ forest data now rule $N_1$ from DW accounting for all of dark matter, requiring some other production mechanism (see \cite{Horiuchi:2013noa} for a summary).\footnote{However, $N_1$ produced through the DW mechanism can still constitute a significant fraction of the dark matter abundance.} Such a mechanism is naturally provided in our framework by the scalar $\phi$. We assume that $\phi$ has additional interactions that keep it in equilibrium with the thermal bath at high temperatures\footnote{Such interactions most likely exist since $\phi$ needs to acquire a vev tied to the SUSY breaking scale, but this assumption is not necessary, and a relic abundance can be obtained even when this does not hold (see \cite{Roland:2014vba}).}; dark matter is then produced through the freeze-in mechanism \cite{freezein}.

\textit{IR freeze-in:} Once $\phi$ obtains a vev, an $N_1$ abundance can be gradually accumulated through the decay channels $\phi\rightarrow N_1\,N_1$ and $H_u\rightarrow N_1 \nu_a$ (with tiny effective couplings $x_{1}=\frac{2\, x\, \pv}{M_*}$ and $y_1=\frac{y\, \pv}{M_*}$ respectively). The abundance due to  $\phi\rightarrow N_1\,N_1$, for instance, is \cite{irabundance}
\begin{equation}
\Omega_{N_1} h^2\sim 0.1 \left(\frac{x_1}{1.4\times10^{-8}}\right)^3 \left(\frac{\pv}{m_{\phi}}\right).
\end{equation}
For $\pv/m_{\phi}\sim\mathcal{O}(1),\,x\sim 1,$ and $\pv\sim 1-100$ PeV, this can be a significant contribution to the dark matter abundance. 

\textit{UV freeze-in:} High temperatures in the early Universe can overcome the 1$/M_*$ suppression of the higher-dimensional interactions from the terms in Eq.\,\ref{eq:newterms}, producing dark matter through the annihilation processes $\phi\,\phi\rightarrow N_1\,N_1$, $\phi\,H_u\rightarrow \nu_a\,N_1$, $\phi\,\nu_a\rightarrow H_u\,N_1$, and $H_u\,\nu_a\rightarrow \phi\,N_1$. The contribution from $\phi\,\phi\rightarrow N_1\,N_1$ to the dark matter relic density, for instance, is approximately \cite{Elahi:2014fsa}
\begin{equation}
\Omega_{N_1} h^2\sim 0.1\,x^2\left(\frac{m_s}{10\,\rm{GeV}}\right)\left(\frac{1000\,T_{RH}\,M_{P}}{M_*^2}\right).
\end{equation}
If the reheat temperature $T_{RH}$ is sufficiently high, such UV freeze-in contributions can also be significant. 

\begin{figure}[htb]
\centering
\includegraphics[height=2.8in]{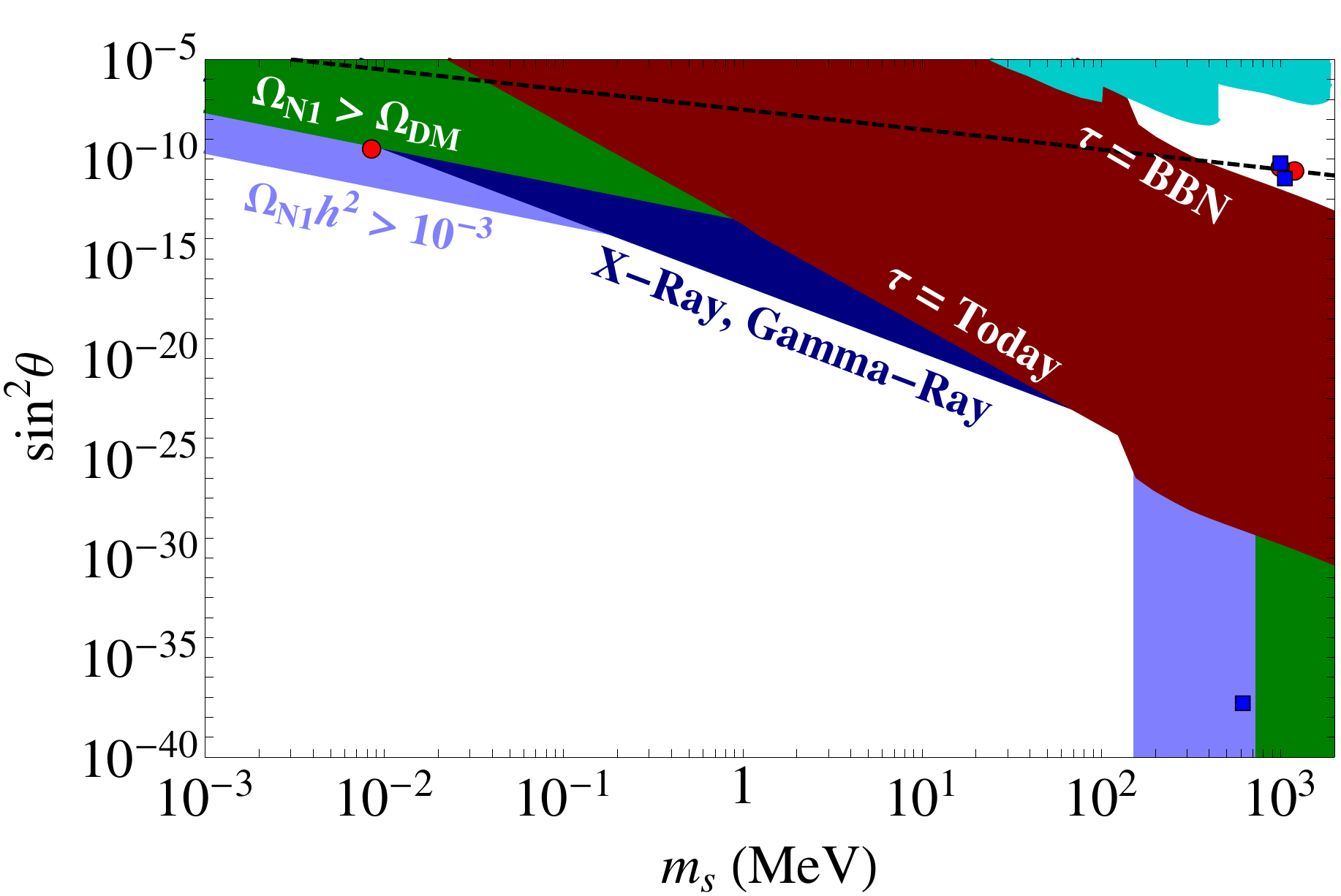}
\caption{Sterile neutrino parameter space. The various colored regions represent various constraints (see \cite{Roland:2014vba} for details). The light blue shaded regions denote parameter space where $10^{-3}\leq \Omega h^2\leq 0.12$; the top left region corresponds to DW production, while the bottom right corresponds to IR freeze-in ($M_*\!=\!M_{GUT}\!=\!10^{16}$ GeV and $\pv=m_{\phi}=100$\,PeV everywhere in the plot).}
\label{fig:constraints}
\end{figure}

Figure \ref{fig:constraints} presents the various masses and mixing angles for $N_1$ for which the correct relic density can be obtained. The light blue shaded regions represent parameter space where $10^{-3}\leq \Omega h^2\leq 0.12$; two distinct regions occur, corresponding to production through the DW mechanism in the top left for keV scale masses and through IR freeze-in at the bottom right (with $\pv=m_{\phi}=100$\,PeV). Other colored regions represent various constraints; the blue squares and red dots represent specific benchmark scenarios in the framework (see \cite{Roland:2014vba} for details). Note that the heavier sterile neutrinos $N_2$ and $N_3$ are generally required to decay before Big Bang Nucleosynthesis (BBN), which forces $\tau_{N2,N3}\leq1$s and consequently $m_{N2,N3}\geq\mathcal{O}(100)$ MeV.

\section{Dark Matter Anomalies}

\subsection{IceCube PeV Neutrinos}

The IceCube collaboration has observed energetic neutrinos at $50$ TeV - $2$ PeV energies, beyond what is expected from astrophysical background, which could come from the decay of a heavy dark matter particle with lifetime $\tau\sim 10^{27}$\,s. Such a dark matter candidate can be accommodated in a straightforward extension of the framework presented here, using the same structure and $U(1)'$  symmetry. In particular, the introduction of two additional superfields $\mathcal{X}$ and $\mathcal{Y}$, analogous to $\mathcal{N}$ and $\Phi$, as defined in Table \ref{table:charges}, leads to the following superpotential:
\begin{eqnarray}
\label{eq:superpot}
{\mathcal{W}}\supset && ~\frac{\zeta_{ij}}{M_*} L_i H_u \,\mathcal{N}_j \Phi+\frac{\alpha_i}{M_*} L_i H_u \mathcal{X} \mathcal{Y}\,+\frac{\eta_i}{M_*}\mathcal{N}_i \mathcal{N}_i \Phi\Phi+\frac{\lambda_1}{M_*}\mathcal{X X Y Y}+\frac{\beta_i}{M_*}\mathcal{N}_i \Phi \mathcal{X Y}\,\nonumber\\
&& ~ +\frac{1}{5!}\frac{\lambda_2}{M_*^3} \mathcal{X} \Phi^5\,+\frac{1}{5!}\frac{\lambda_3}{M_*^3} \mathcal{Y N}_i^5.
\end{eqnarray}

\begin{table} [h]
\begin{center}
\begin{tabular}{ | c | c | c | c | l |}
\hline
~Supermultiplet~ & ~spin 0, 1/2~ & $~U(1)'~$ &~ Remarks \\
\hline
  $\mathcal{N}_i$ & $\tilde{N_i}$, $N_i$ & +1 &  ~$N_i$ sterile neutrinos ~\\
  $\Phi$ & $\phi, \psi_\phi$ & -1 &  ~$\langle\phi\rangle\sim$PeV, breaks $U(1)'$ ~\\
  $\mathcal{X}$ & X, $\psi_X$ & +5 &  ~$m_X\sim$PeV, dark matter~ \\
  $\mathcal{Y}$ & Y, $\psi_Y$ & -5 &  ~ $U(1)'$ partner of $\mathcal{X}$~\\
  \hline
\end{tabular}
\caption{Field content, notation, and $U(1)'$ charge assignments for the new multiplets.}
\label{table:charges}
\end{center}
\end{table}

The SM singlet scalar $X$ is our dark matter candidate that produces the PeV neutrinos observed by IceCube; it obtains a PeV scale mass after $U(1)'$ and SUSY breaking, and its large $U(1)'$ charge makes it sufficiently long lived. 

The lowest dimension term connecting $X$ with the SM fields is $\frac{\alpha_i}{M_*}L_i H_u \mathcal{X} \mathcal{Y}$ (see Eq.\,\ref{eq:superpot}), which results in the following feeble (freeze-in) production processes for $X$ from the thermal bath:
\begin{equation}
l\,h\,\rightarrow X \,\psi_Y,~~~~~ l\,\tilde{H}\,\rightarrow X \,Y,~~~~~\tilde{l}\,\tilde{H}\,\rightarrow X\,\psi_Y.
\label{xproduction}
\end{equation}
Here $l$ denotes both charged leptons and neutrinos, and $h$ denotes both neutral and charged higgses, and likewise for their superpartners $\tilde{l}$ and $\tilde{H}$. The same processes also result in freeze-in abundances of $Y,\, \psi_X$, and $\psi_Y$, but these eventually  decay into $X$, which is assumed to be lighter. Taking all these contributions into account, the relic abundance of $X$ is calculated to be approximately 
\begin{equation}
\Omega_{X}h^2 \sim 0.12 ~\bigg(\frac{m_{X}}{10 ~\rm{PeV}}\bigg) \bigg(\frac{\alpha}{10^{-5}} \bigg)^2  \left(\frac{T_{RH}}{1.5 \times 10^{9} \,\rm{GeV}} \right) \left(\frac{10^{16} ~\rm{GeV}}{M_*} \right)^2
\end{equation}
where we have taken $\alpha=\alpha_i$ for simplicity. Therefore, with a sufficiently high reheat temperature $T_{RH}$ and appropriate values of $\alpha$, the PeV scale particle $X$ could compose a significant fraction of dark matter.

Decays of $X$ arise from the superpotential term $\frac{1}{5!}\frac{\lambda_2}{M_*^3} \mathcal{X} \Phi^5\,$. Assuming $\langle \phi\rangle\textgreater \,m_\phi$, the leading decay process is $X\rightarrow \psi_\phi\,\psi_\phi$, which (assuming $m_{\psi_\phi}/m_X\ll 1$) results in a lifetime
\begin{equation}
\label{eq:lifetime}
\tau_X\approx 10^{27} \,s\, \left(\frac{1.5\times10^{-3}}{\lambda_2}\right)^2\,\left(\frac{M_*}{10^{16}\, \rm{GeV}} \right)^6 \, \left(\frac{100\, \rm{PeV}}{\pv} \right)^6 \, \left(\frac{\rm{PeV}}{m_X} \right).
\end{equation}

\begin{figure}[t]
\centering
\includegraphics[height=2.8in]{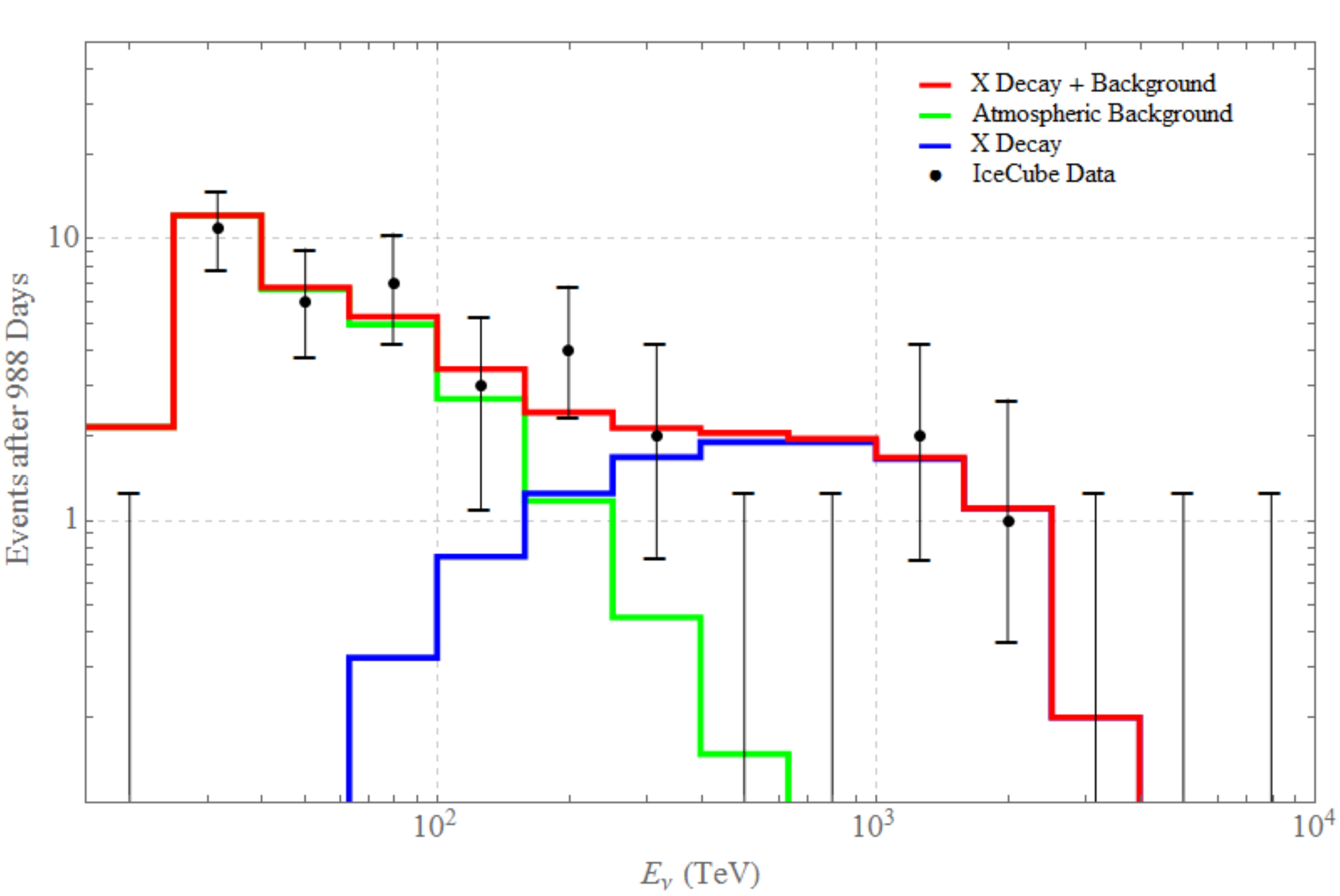}
\caption{Expected number of neutrino events from decays of $X$ (blue), IceCube data points with error bars (black), atmospheric background (green), and total signal+background events (red).}
\label{fig:decayspectrum}
\end{figure}

The $\psi_\phi$ further decays as $\psi_\phi\rightarrow N\tilde{H}\nu\,,\psi_\phi\rightarrow N\tilde{H^\pm}l^\mp$ through an off-shell sterile sneutrino as a consequence of the $L_i H_u \,\mathcal{N}_j \Phi$ and $\mathcal{N}_i \mathcal{N}_i \Phi\Phi$ terms in the superpotential. The sterile neutrinos $N$ then further decay through the standard sterile neutrino decay channels to produce additional active neutrinos. The decay lifetime required to fit the IceCube data is $\tau\sim 10^{27}$\,s; Eq.\,\ref{eq:lifetime} suggests that one can obtain the necessary lifetime for reasonable choices of parameters in the model. Fig. \ref{fig:decayspectrum} shows the number of neutrinos expected at IceCube from the decays of $X$ for such parameter choices (see \cite{Roland:2015yoa} for details); note that the multiple step decay chain into neutrinos results in a flat shape in agreement with the IceCube data. 

\subsection {3.5 keV X-Ray Line}

Several papers have interpreted this signal as the decay of a $\sim\!7$ keV sterile neutrino. In the framework presented here, such a sterile neutrino can be produced through DW and form only a fraction of dark matter, evading the combined constants from X-rays and Lyman-$\alpha$. In a warm plus cold dark matter setup, \cite{Harada:2014lma} found that the 3.5 keV X-ray signal could be explained with a 7 keV sterile neutrino produced from DW that made up $10-60 \%$ of dark matter, and the parameters in our framework can be chosen to map on to one of the benchmark points presented in this paper: a $7$ keV sterile neutrino with mixing angle sin$^2 (2\theta)\sim 4\times 10^{-10}$, making up $\sim 25 \%$ of dark matter. Hence the $3.5$ keV X-ray line signal can be incorporated in our framework. Note that such a mixture of warm and cold components for dark matter could offer several advantages. 

\section{Summary}
In summary, we have presented an extended supersymmetric neutrino sector with the right-handed neutrinos charged under an additional symmetry ($U(1)'$) that is broken by a PeV scale vev. This framework features:
\begin{itemize}
\item Active neutrino masses in agreement with oscillation data (achieved with predominantly $\mathcal{O}(1)$ couplings in the theory)
\item keV-GeV scale sterile neutrinos, with freeze-in production of sterile neutrino dark matter
\item connections to PeV scale supersymmetry
\item PeV energy neutrinos at IceCube from the decay of a scalar dark matter component $X$, from a straightforward extension of the framework. 
\item the 3.5 keV X-ray line from the decay of a 7 keV sterile neutrino
\end{itemize}

\end{document}

%% file: econfmacros.tex
\def\beq{\begin{equation}}
\def\eeq#1{\label{#1}\end{equation}}
\def\eeqn{\end{equation}}

\def\beqa{\begin{eqnarray}}
\def\eeqa#1{\label{#1}\end{eqnarray}}
\def\eeqan{\end{eqnarray}}

\let\bar=\overbar

\def\Dslash{\not{\hbox{\kern-4pt $D$}}}
\def\dslash{\not{\hbox{\kern-2pt $\del$}}}

\def\msb{{\bar{\ssstyle M \kern -1pt S}}}

%% file: eprint_dpf2015.bbl
\begin{thebibliography}{99}

%%
%%  bibliographic items can be constructed using the LaTeX format in SPIRES:
%%    see    http://www.slac.stanford.edu/spires/hep/latex.html
%%  SPIRES will also supply the CITATION line information; please include it.
%%

\bibitem{numsm} 
  T.~Asaka, S.~Blanchet and M.~Shaposhnikov,
  %``The nuMSM, dark matter and neutrino masses,''
  Phys.\ Lett.\ B {\bf 631}, 151 (2005)
  [hep-ph/0503065].
  T.~Asaka and M.~Shaposhnikov,
  %``The nuMSM, dark matter and baryon asymmetry of the universe,''
  Phys.\ Lett.\ B {\bf 620}, 17 (2005)
  [hep-ph/0505013].
  T.~Asaka, M.~Shaposhnikov and A.~Kusenko,
  %``Opening a new window for warm dark matter,''
  Phys.\ Lett.\ B {\bf 638}, 401 (2006)
  [hep-ph/0602150].
  T.~Asaka, M.~Laine and M.~Shaposhnikov,
  %``Lightest sterile neutrino abundance within the nuMSM,''
  JHEP {\bf 0701}, 091 (2007)
  [JHEP {\bf 1502}, 028 (2015)]
  [hep-ph/0612182].
  
  \bibitem{DW} 
  S.~Dodelson and L.~M.~Widrow,
  %``Sterile-neutrinos as dark matter,''
  Phys.\ Rev.\ Lett.\  {\bf 72}, 17 (1994)
  [hep-ph/9303287].
  
  \bibitem{35kev}
  E.~Bulbul, M.~Markevitch, A.~Foster, R.~K.~Smith, M.~Loewenstein and S.~W.~Randall,
  %``Detection of An Unidentified Emission Line in the Stacked X-ray spectrum of Galaxy Clusters,''
  Astrophys.\ J.\  {\bf 789}, 13 (2014)
  [arXiv:1402.2301 [astro-ph.CO]].
  A.~Boyarsky, O.~Ruchayskiy, D.~Iakubovskyi and J.~Franse,
  %``Unidentified Line in X-Ray Spectra of the Andromeda Galaxy and Perseus Galaxy Cluster,''
  Phys.\ Rev.\ Lett.\  {\bf 113}, 251301 (2014)
  [arXiv:1402.4119 [astro-ph.CO]].

\bibitem{Aartsen:2014gkd} 
  M.~G.~Aartsen {\it et al.} [IceCube Collaboration],
  %``Observation of High-Energy Astrophysical Neutrinos in Three Years of IceCube Data,''
  Phys.\ Rev.\ Lett.\  {\bf 113}, 101101 (2014)
  [arXiv:1405.5303 [astro-ph.HE]].

\bibitem{Roland:2015yoa} 
  S.~B.~Roland, B.~Shakya and J.~D.~Wells,
  %``PeV Neutrinos and a 3.5 keV X-Ray Line from a PeV Scale Supersymmetric Neutrino Sector,''
  arXiv:1506.08195 [hep-ph].
  %%CITATION = ARXIV:1506.08195;%%
  
  \bibitem{Roland:2014vba} 
  S.~B.~Roland, B.~Shakya and J.~D.~Wells,
  %``Neutrino Masses and Sterile Neutrino Dark Matter from the PeV Scale,''
  arXiv:1412.4791 [hep-ph].
  %%CITATION = ARXIV:1412.4791;%%
  
  \bibitem{pevsusy}
  J.~D.~Wells,
  %``Implications of supersymmetry breaking with a little hierarchy between gauginos and scalars,''
  hep-ph/0306127. 
   N.~Arkani-Hamed and S.~Dimopoulos,
  %``Supersymmetric unification without low energy supersymmetry and signatures for fine-tuning at the LHC,''
  JHEP {\bf 0506}, 073 (2005)
  [hep-th/0405159].  
  G.~F.~Giudice and A.~Romanino,
  %``Split supersymmetry,''
  Nucl.\ Phys.\ B {\bf 699}, 65 (2004)
  [Nucl.\ Phys.\ B {\bf 706}, 65 (2005)]
  [hep-ph/0406088].  
    J.~D.~Wells,
  %``PeV-scale supersymmetry,''
  Phys.\ Rev.\ D {\bf 71}, 015013 (2005)
  [hep-ph/0411041].
  
  \bibitem{Horiuchi:2013noa} 
  S.~Horiuchi, P.~J.~Humphrey, J.~Onorbe, K.~N.~Abazajian, M.~Kaplinghat and S.~Garrison-Kimmel,
  %``Sterile neutrino dark matter bounds from galaxies of the Local Group,''
  Phys.\ Rev.\ D {\bf 89}, no. 2, 025017 (2014)
  [arXiv:1311.0282 [astro-ph.CO]].
  
  \bibitem{freezein}
    L.~J.~Hall, K.~Jedamzik, J.~March-Russell and S.~M.~West,
  %``Freeze-In Production of FIMP Dark Matter,''
  JHEP {\bf 1003}, 080 (2010)
  [arXiv:0911.1120 [hep-ph]].
  
  \bibitem{irabundance}
  A.~Kusenko,
  %``Sterile neutrinos, dark matter, and the pulsar velocities in models with a Higgs singlet,''
  Phys.\ Rev.\ Lett.\  {\bf 97}, 241301 (2006)
  [hep-ph/0609081].
  K.~Petraki and A.~Kusenko,
  %``Dark-matter sterile neutrinos in models with a gauge singlet in the Higgs sector,''
  Phys.\ Rev.\ D {\bf 77}, 065014 (2008)
  [arXiv:0711.4646 [hep-ph]].
  
  \bibitem{Elahi:2014fsa} 
  F.~Elahi, C.~Kolda and J.~Unwin,
  %``UltraViolet Freeze-in,''
  JHEP {\bf 1503}, 048 (2015)
  [arXiv:1410.6157 [hep-ph]].
  
   \bibitem{Harada:2014lma} 
  A.~Harada, A.~Kamada and N.~Yoshida,
  %``Structure formation in a mixed dark matter model with decaying sterile neutrino: the 3.5 keV X-ray line and the Galactic substructure,''
  arXiv:1412.1592 [astro-ph.CO].


\end{thebibliography}
